\documentclass[10pt,letterpaper]{article}

\usepackage[top=0.85in,left=2.75in,footskip=0.75in]{geometry}

\usepackage{amsmath,amssymb}

\usepackage{changepage}

\usepackage[utf8x]{inputenc}

\usepackage{textcomp,marvosym}

\usepackage{cite}

\usepackage{nameref,hyperref}

\usepackage[right]{lineno}

\usepackage{microtype}
\DisableLigatures[f]{encoding = *, family = * }

\usepackage[table]{xcolor}

\usepackage{array}

\usepackage{tabularx}
\usepackage{bm}
\usepackage{multirow}
\usepackage{booktabs}
\usepackage{rotating}
\usepackage{ragged2e}
\usepackage{mathtools}
\usepackage{enumitem}
\usepackage{pdflscape}
\usepackage{afterpage}

\newcommand{\Signalling}{{\bf S}}
\newcommand{\SingleCell}{{\bf SCB}}
\newcommand{\CollectiveCell}{{\bf CCB}}

\newcolumntype{b}{X}
\newcolumntype{s}{>{\hsize=.5\hsize}X}
\newcolumntype{t}{>{\hsize=.25\hsize}X}
\newcolumntype{Y}{>{\RaggedRight\arraybackslash}X}
\newcolumntype{+}{!{\vrule width 2pt}}

\newlength\savedwidth

\newcommand{\bit}{\begin{itemize}}
\newcommand{\eit}{\end{itemize}}
\newcommand{\beq}{\begin{equation}}
\newcommand{\eeq}{\end{equation}}

\definecolor{andreascolor}{rgb}{0.30,0.03,0.66}

\newcommand{\Andreas}[1]{#1}
\newcommand{\LEK}[1]{#1}

\graphicspath{{./Figures/}{./include/msc_figs/}{./include/constvol/}}
\makeatletter
\def\input@path{{./Figures/}{./include/msc_figs/}}
\makeatother

\raggedright
\usepackage[aboveskip=1pt,labelfont=bf,labelsep=period,justification=raggedright,singlelinecheck=off]{caption}

\makeatletter
\renewcommand{\@biblabel}[1]{\quad#1.}
\makeatother

\usepackage{lastpage,fancyhdr,graphicx}
\usepackage{epstopdf}
\fancyheadoffset[L]{2.25in}
\fancyfootoffset[L]{2.25in}
\begin{document}
\vspace*{0.2in}

\begin{flushleft}
{\Large
\textbf\newline{Bridging from single to collective cell migration: A review of models and links to experiments.}
% Please use "sentence case" for title and headings (capitalize only the first word in a title (or heading), the first word in a subtitle (or subheading), and any proper nouns).
}
\newline
\\
Andreas Buttenschön\textsuperscript{1, *},
Leah Edelstein-Keshet\textsuperscript{1}
\\
\bigskip
\textbf{1} Department of Mathematics, University of British Columbia, Vancouver, Canada \\
\bigskip

% Insert additional author notes using the symbols described below. Insert symbol callouts after author names as necessary.
%
% Remove or comment out the author notes below if they aren't used.
%
% Primary Equal Contribution Note
% \Yinyang These authors contributed equally to this work.

% Additional Equal Contribution Note
% Also use this double-dagger symbol for special authorship notes, such as senior authorship.
% \ddag These authors also contributed equally to this work.

% \textcurrency Current Address: Dept/Program/Center, Institution Name, City, State, Country % change symbol to "\textcurrency a" if more than one current address note
% \textcurrency b Insert second current address
% \textcurrency c Insert third current address

% Group/Consortium Author Note
% \textpilcrow Membership list can be found in the Acknowledgments section.

% Use the asterisk to denote corresponding authorship and provide email address in note below.
* andreas@buttenschoen.ca

\end{flushleft}
\section*{Abstract}
Mathematical and computational models can assist in gaining an understanding of
cell behavior at many levels of organization. Here, we review models in the
literature that focus on eukaryotic cell motility at 3 \Andreas{size scales}:
intracellular signaling that regulates cell shape and movement, single cell
motility, and collective cell behavior from a few cells to tissues. We survey
recent literature to summarize distinct computational methods (phase-field,
polygonal, Cellular Potts, and spherical cells). We discuss models that bridge
between levels of organization, and describe levels of detail, both biochemical
and geometric, included in the models. We also highlight links between models
and experiments. We find that models that span the 3 levels are still in the
minority.

% Please keep the Author Summary between 150 and 200 words
% Use first person. PLOS ONE authors please skip this step.
% Author Summary not valid for PLOS ONE submissions.
\section*{Author summary}

In this review paper, we summarize the literature on computational models for
cell motility, from the biochemical networks that regulate it, to the behavior
of 1 and many cells. We discuss the distinct approaches used at each level,
and how models can build bridges between the \Andreas{different size scales.}
We find models at many different levels of biological detail, and discuss their
relative contributions to our understanding of single and collective cell
behavior. Finally, we indicate how models have been linked to biological
experiments in this field.

% \linenumbers

\section{Introduction}

Over several decades, there has been great progress in our understanding of cell
motility. In the 1980s and 1990s, the basic machinery of eukaryotic cell
motion and the role of the actin cytoskeleton were discovered and refined.
Regulation of motility by intracellular signaling networks was then deciphered
in the late 1990s and through the 2000s. We continue to discover links between
cell signaling and cell shape and function, in both normal and diseased cells.
Recent efforts aim to link single cell behavior to collective behavior of many
cells and emergent dynamics of tissues.

Though originally descriptive, cell biology has emerged as a quantitative
science over the same time span. Mathematical and computational modeling have
become more universally accepted, more closely integrated with experimental
research, and more advanced in terms of methodology.

Here, we survey the state of the field, emphasizing bridges that span scales:
from molecular signaling to multicellular hierarchies. We focus on the role of
modeling and computational biology. Because the literature is vast and growing
exponentially, we limit our review to several key themes and concentrate on
3 questions:

\begin{enumerate}[label=\textbf{(\arabic*)},ref=(\arabic*),leftmargin=*,labelindent=\parindent]
    \item To what extent have models provided a way to bridge between the 3 levels of organization,
    from intracellular signaling, to single cell behavior, and to collective
    cell/tissue behavior?
    \item What level of detail is appropriate in a computational or mathematical model? What kinds of models are suitable for a given situation?
    \item What is the relationship between models and experiments in the current literature on the subject?
\end{enumerate}
\LEK{At each level, we consider these 3 questions in subsections with headings
``Bridging scales,'' ``Levels of detail,'' and ``Links with experiments.'' Like
any other subdivision, this is to some extent arbitrary, as literature papers
often span such categories. }

\begin{figure}
 \centering
 \includegraphics[width=\textwidth]{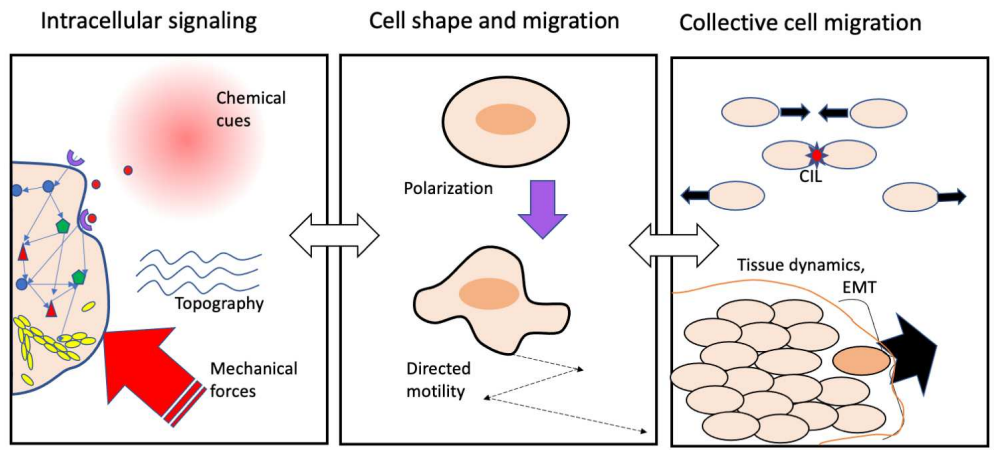}
 \captionof{figure}{Mathematical models can be used to bridge from intracellular
    signaling (left), to single cell shape and motility (center), to cell--cell
    interactions (right). At the lowest scales, the goal is deciphering the
    interplay between stimuli to the cell (chemical, topographic, mechanical,
    etc.) and intracellular signaling networks that regulate F-actin (branched
    polymer) and the cytoskeleton (not drawn to scale). These interactions lead to
    protrusion or retraction, cell polarization, and shape changes that enable
    directed motility and chemotaxis. At a higher level, an aim is to link cell
    behavior and cell--cell interactions to the outcomes of cell collisions
    (e.g., CIL) and to the cohesion of tissues versus EMT, where cells break off. Interconnections
    exist between all layers, only 2 of which (white arrows) are shown here.
    CIL, Contact Inhibition of Locomotion; EMT, Epithelial Mesenchymal
    Transition. }
   \label{fig:ScaleUp}
\end{figure}

Many excellent reviews are already available, including
\cite{blanchard2018,Spatarelu2019,Sun2017,Alert2019}.
Some survey computational methods and others provide links to experiments. The
focus on the above set of 3 questions is, to our knowledge, unique to the
current review.

\begin{figure}[!h]
 \centering
 \includegraphics[width=\textwidth]{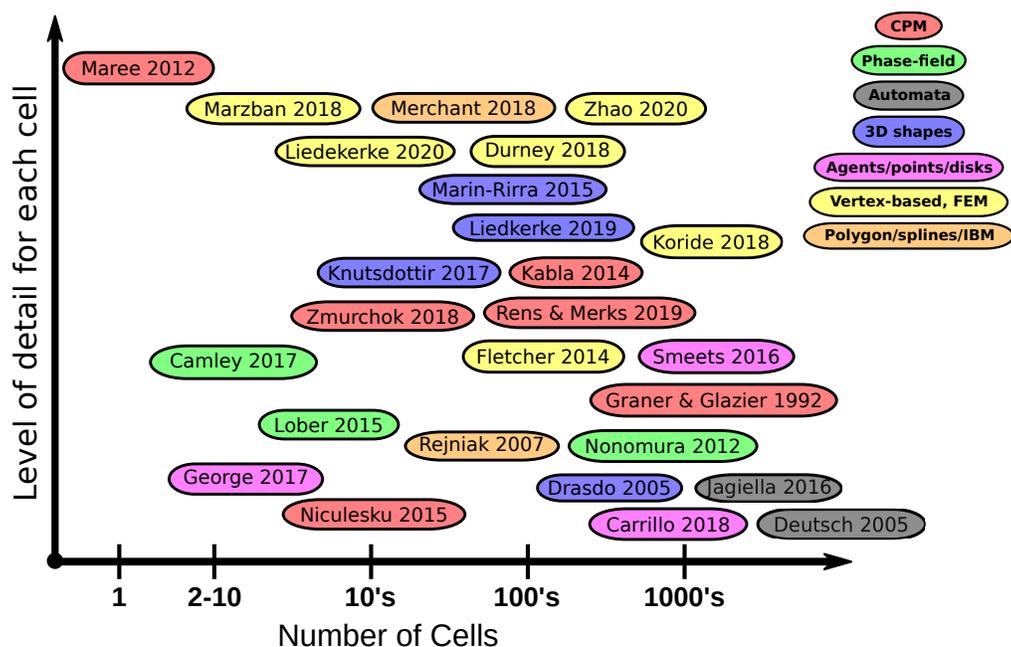}
 \captionof{figure}{A mapping of computational models according to the number of cells
     (horizontal axis) and the level of detail for each cell (vertical axis).
     Citations of papers in the diagram \Andreas{(starting from the upper left
     to lower right}: \cite{Maree2012,Marzban2018a,Merchant2018,Zhao2020,Liedekerke2020,Durney2018,Marin2015,vanLiedekerke2019b,Koride2018,Knutsdottir2017,Kabla2012,Zmurchok2018,Rens2019b,Camley2017,Fletcher2014,Smeets2016,Graner1992,Lober2015,Rejniak2007,Nonomura2012,George2017,Niculescu2015,Drasdo2005,Jagiella2016,Carrillo2018,Deutsch2005}).
    }\label{fig:ModelMap}
\end{figure}

The paper is organized by size-scale and level of detail. As shown in
Fig~\ref{fig:ScaleUp}, we start with the subcellular level of  biochemical
signaling (left), and move up to single cell behavior (center). We then link to
small cell groups, larger groups, and tissues (right). At each level, we revisit
the 3 key themes and select a few representative  contributions from the
literature to use as examples. A summary ``mapping'' of the modeling literature
into levels of detail and numbers of cells is provided in
Fig~\ref{fig:ModelMap}.

\Andreas{%
No one review paper can do justice to the entire field. Hence, we point the
reader to related review articles that complement our own. In some cases, they
cover similar ground but with distinct emphases or points of view.
In \cite{blanchard2018}, the authors study the issue of cell heterogeneity, its
sources at various size scales, and its role in collective cell migration. They
briefly discuss self-propelled particles (SPP), Cellular Potts Models (CPM), and
vertex-based approaches that we also discuss in this review. They recommend
further investigation of mechanical assumptions in models, and of more realistic
tissue size.  A thorough review of the biomechanics of collective cell migration
appears in \cite{Spatarelu2019}, with a primary focus on cancer.  In contrast
with other reviews, this paper also provides an excellent summary of
experimental methods.  Phenomena discussed include the ``unjamming transition''
where a cell collective changes from ``solid-like'' to ``liquid-like'' behavior
and its connection to the epithelia-mesenchymal transition.}

\Andreas{%
Force balance and energy-based models for single amoeboid and collective
mesenchymal cell migration are compared in \cite{Sun2017}. The authors indicate
the challenge of converting between energy-based and force-based modeling
platforms, and point to the significance of doing so. (See \cite{Rens2019} for
this connection in the CPM.) They also describe instances of close
experiment-model integration.  Finally, a recent review of multiscale models in
\cite{Alert2019} has a soft-matter physics perspective and describes
computational methods (CPM, phase-field, active matter, particle systems, etc.)
and their physical basis.
}

\section{From cytoskeleton and intracellular signaling to cell shape and migration}

How do chemical and mechanical stimuli, together with intracellular signaling
shape the behavior of single cells?  This question is central to the bridge
between left and central panels of Fig~\ref{fig:ScaleUp}.

\subsection{Actin dynamics}

We briefly highlight this well-established area to illustrate examples of
bridging scales, diverse levels of detail, and model--experiment synergy
\cite{Pollard2001}.

\subsubsection{Bridging scales}
Actin dynamics is studied at many levels: from
association of actin monomers to form filaments \cite{Mogilner2002}, to
biophysical force production by F-actin  \cite{Mogilner1996}, the assembly and
branching of F-actin \cite{Grimm2003}, and the resultant shape and motion of
cells \cite{Lacayo2007,Keren2008}.

An excellent review of actin dynamics models in single cell migration,
\cite{Danuser2013} provides a solid bridge between molecular scale and cell
scale phenomena. The authors explain the main model formalisms and show how
models can be used to understand experimental data.  A review of the link from
actin and its properties to cell mechanosensing and behavior is given in
\cite{Mak2015}. \LEK{Reviews of the literature on actin-based cell migration reveal a mature field, where quantitative methods and theory have ripened in tandem. This synergy has benefited greatly from biochemical pioneers, such as Tom Pollard (Yale University),  who helped to nurture an appreciation for mathematical and physical modeling in the community.}

\subsubsection{Levels of detail}
The pair of papers \cite{Ditlev2009,Mogilner2002} aptly
illustrates the dichotomy between highly detailed computational models
\cite{Ditlev2009} and more conceptual minimal models \cite{Mogilner2002}.
On one hand, the highly detailed \cite{Ditlev2009} synthesizes a large body of
experimental data for actin assembly and branching, including many actin-related
component. At the opposite extreme, the models of \cite{Mogilner2002}, as well
as later papers \cite{Keren2008,Lacayo2007}
emphasize physical principles, universal properties, and general insights. In
some sense, the overall
predictions of both model types can be compared. The minimal models are more
tractable for analysis, parameter sweeps, and overall insights, but are harder
to connect to detailed molecular biology experiments.
\Andreas{The complexity of highly detailed computational models (matching
molecular experiments) makes it harder to navigate their results, which are more
like a ``1-to-1'' map.} Such maps may be important for those
who ``live in the neighborhood,'' but are essentially baffling for newcomers.
While distinct, one could argue that these approaches are complementary, so that
overall insights can be obtained from one, and specific details from the other.
\LEK{The dichotomy between the detailed and the simplified models will reappear
throughout this review, serving as 1 hallmark of distinct views.}

\subsubsection{Links with experiments}
A review of the experiments and theory for actin
dynamics in the motility of keratocytes is provided in \cite{Mogilner2019}. A
retrospective that emphasizes the importance of theory, and case-study where
theory has led the way is described in \cite{Pollard2016}.  Notable
contributions linking experiments to theory for cell shape include
\cite{Lacayo2007,Keren2008}, where shapes of cells are classified and related to actin
branching and treadmilling.

\subsection{Signaling networks}

Fluorescence resonance energy transfer (FRET) microscopy led to breakthroughs in
visualizing the activities of signaling proteins that regulate the cytoskeleton.
This has enabled live imaging of spatiotemporal activity of the Rho family
GTPases \cite{Sekar2003}.

The roles of Rho family GTPases (Cdc42, Rac, and Rho) in regulating actin
assembly and myosin contraction are well--known \cite{Hall1998,Hall2000}.
\LEK{These switch-like proteins are central regulators of cell migration,
funneling signals from the cell's environment to downstream components that
shape its motility.} Cdc42 and Rac promote F-actin assembly and cell edge
protrusion, whereas Rho facilitates myosin light chain (MLC) phosphorylation,
activating myosin-based cell contraction. Rho also promotes F-actin through
formins and is a Rac antagonist.  The importance of
GTPase dynamics in cell motility \cite{Ridley2001} has also attracted modelers
to this arena.

A review of GTPase spatio temporal models can be found in \cite{Khatibi2018}. A
classic paper on modeling spatio temporal dynamics of signaling in cells (not
necessarily GTPases) is \cite{Kholodenko2006}, a paper that emphasizes
universal principles. We see how commonly shared basic motifs combine to form
complex dynamics \cite{Kholodenko2006}. Many lessons learned from this approach
can be applied directly to studying GTPases or other signaling networks. \LEK{In
general, the community stands to benefit from more expository papers of this
type, where the ingredients that combine to set up specific dynamical signatures
are exposed. See also \cite{Tyson2003} for a popular example of this type, based
on years of experience with models of cell cycle regulatory networks. }

\subsubsection{Bridging scales}

Several papers link the dynamics of GTPases to cytoskeletal assembly and cell
shape. Among these are experimental \cite{Sailem2014,Byrne2016} and
computational \cite{Maree2006,Maree2012} works. A review of signaling that
organizes the cell  front and rear in Dictyostelium discoideum is given in
\cite{Devreotes2017}, who also survey models of these as excitable systems.
Some of these papers are described in fuller detail below.

\subsubsection{Levels of detail}

GTPases act like molecular switches that cycle on and off the cell membrane and
interact via crosstalk through effector proteins. \LEK{A large collection of
proteins participate in GTPase signaling: Guanine nucleotide exchange factors
(GEFs) activate and GTPase-activating proteins (GAPs) inactivate the GTPases
with varying degrees of specificity.  GTPases are also sequestered in the
cytosol by binding to guanine nucleotide dissociation inhibitors (GDIs).} In
some models,
notably \cite{Otsuji2007,Jilkine2007}, the details of the binding and mechanisms
of activation--inactivation are lumped into phenomenological terms such as
Michaelis--Menten or Hill functions. In \cite{Jilkine2007}, emphasis on
cross-talk of Cdc42, Rac, and Rho and on spatial polarization comes at the
expense of the molecular steps themselves (some of which remain unknown). In
\cite{Sakumura2005}, more detail on such steps is modeled and eventually reduced
to 3 ordinary differential equations (ODEs) using a quasi-steady-state approach.
The authors focus on the role of a Cdc42-GEF in oscillatory phenotype in
neuronal growth cone motility. In
contrast, \cite{Falkenberg2013} concentrate on GDI binding of GTPases in a
highly detailed computational model (originally crafted in BioNetGen and Virtual
Cell).

Crosstalk between GTPases is modeled in great detail by \cite{Hetmanski2016},
who constructed a Boolean model for epidermal growth factor (EGF)
signaling to Rac and Rho. Their model
has 38 intermediates and multiple reactions that activate or inhibit
components. (See also \cite{Letort2018}, who combines PhysiCell with MaBoss to
allow the modeling of intracellular signaling networks as Boolean networks.)
At lower level of detail are Rac-Rho mutually antagonistic models that leave out
the forest of interacting nodes \cite{Holmes2016}.

Simpler models for single
GTPase include \cite{Mori2008} for cell polarity and
\cite{Vanderlei2011,Cusseddu2019} for cell shape. There, assumptions are made to
condense underlying complex interactions into simpler ``rules'' of behavior.
For example, ``high GTPase activity leads to outwards protrusion force at the
cell edge'' (in the case of Rac) or ``contraction'' (in the case of Rho)
\cite{Park2017,Holmes2017}.
An advantage of this simplification is that simple models consisting of few partial
differential equations (PDEs) allow for well-honed methods of applied
mathematics, including nonlinear dynamics, stability theory, asymptotic
analysis, and/or bifurcation theory to explore model predictions, parameter
dependence, and regimes of behavior.

\LEK{Serendipitously, simplified signaling models, like \cite{Mori2008},
occasionally also expose interesting mathematical structures to study. A case in
point is ``wave-pinning'' \cite{Mori2008,Mori2011}: a wave of GTPase activity
initiated at 1 end of a cell stalls before reaching the opposite end, resulting
in robust ``cell polarization.''} Simplified models also permit numerical
implementation in more complex
geometries. For example, in \cite{Cusseddu2018}, a wave-pinning model for a
single GTPase in a single 3D cell is implemented by bulk diffusion methods. The
cell does not deform, but the spatial localization of the GTPase is quantified
in a fully 3D geometry.  \Andreas{(See also \cite{Wu2015}, who considered
neutrophil fluidization}, and \cite{Diegmiller2018}, who considered
polarization in a 3D sphere.)

At a gradually increasing scale are  models that include not just GTPases (Rac,
Rho, and Cdc42) but also other layers (phosphoinositides, actin, Arp2/3, and myosin)
that interact to regulate cell polarity in response to stimuli \cite{Lin2012},
or dictate cell shape and directed motility \cite{Maree2006, Maree2012}. In the
latter 2 papers, a single moving ``keratocyte'' is represented using a
Cellular Potts Model (CPM), \LEK{an energy minimization agent-based platform for
computing cell shapes}. It is shown that GTPase signaling can account for cell
polarization, reorientation \cite{Maree2006}, and resolution of conflicting
cues or obstacles \cite{Maree2012}.  (Compare with \cite{Camley2013}, who employ
phase-field computational methods toward similar goals.)

By way of comparison, in another more detailed approach, in \cite{Kopfer2018}, a
cell is modeled as having a solid boundary that is moved using a Stefan
condition. For the chemical signaling, Rho, Rac, 2 species of GEFs, F-actin,
and G-actin are included. The model accounts for the basic repertoire of
neutrophil motility.

\LEK{All in all, models of GTPase signaling have taught us several valuable
lessons, some with universal ramifications. First, modeling has provided clues
to the functional significance of the seemly strange cycling of GTPases between
membrane and cytosol: namely, the separation of diffusion rates resulting from
these distinct compartments could be playing a role in pattern formation
processes, an ingredient enabling GTPase localization or patterning in cells.
Second, the stripped-down models have shown that chemical polarization in a cell
need not depend on crosstalk between multiple types of GTPases --- it can be set
up by a single member of the family, given sufficient positive feedback and some
depletion of its cellular pool \cite{Mori2008}. Groups are still occasionally
rediscovering on their own, the link between cell size and cell polarization
that was implicit in \cite{Mori2011}, suggesting the need for more expository
reviews of mathematical results. An important concept introduced in
\cite{Maree2012}, but not yet fully recognized in the community, is the synergy
between GTPase dynamics and its effect of cell shape and boundary curvature of
the cell edge. Simply put, the ``motion by curvature'' of the chemical system
interacts with boundary conditions to accelerate the dynamics. Altogether, the
appreciation of GTPase signaling has benefited greatly from a host of distinct
modeling and mathematical approaches. }

\subsubsection{Links with experiment}

A review of the links between models of cell migration and experimental data
(image processing, cell tracking, and feature extraction from 1 cell to many) is
given in \cite{Masuzzo2016}. Here we focus more specifically on experiments that
highlight intracellular signaling.

The work by \cite{Lin2012} provides data for the reorientation and polarity
responses of HeLa cells from various starting cell states (polar, anti-polar, or
nonpolar). The authors showed that an internal circuit of signaling (Cdc42, Rac,
Rho, and phosphoinositides) could account qualitatively for the observed
responses of these cells in microfluidic channels, with an externally
controllable response to Rac. In both this and the follow-up \cite{Lin2015}, the
1D geometry of the channels helps to reduce the geometric complexity of cell
shape, allowing for a better match to 1D spatial model representations.

How are models for GTPase crosstalk experimentally linked to cell morphology and
motility dynamics? In experimental results of \cite{Byrne2016}, the authors
showed that signaling circuit of mutually antagonistic Rac and Rho could affect
not only the dynamics of F-actin, but also shapes and migration of mesenchymal
breast cancer cells. Together with a mathematical model for the Rac-Rho
interactions, they were able to manipulate the Rac-Rho competition, demonstrate
bistable states, and show that manipulating the system by inhibiting PAK (\LEK{a kinase that mediates} inhibition of Rho by Rac) displayed hysteresis characteristic of
bistable systems. The significance of this paper is that it demonstrates a
direct link between a simple hypothetical model for the way that Rac and Rho
GTPases operate in a cell and the next level up, that of overall cell
morphology.

Sometimes, individual papers provide only part of the story, but taken as a
whole, a series of papers gives a broader view. The sequence of work in
\cite{Mackay2014,Sailem2014,Holmes2016,Zmurchok2020} explore how Rac-Rho mutual
antagonism is linked to cell morphology.  In \cite{Mackay2014}, the authors experimentally
manipulated Rac and Rho activities to show spread or contracted cells.
Interestingly, they found that combining constitutively active (CA) Rac and Rho
simultaneously results in mixed morphologies in human glioma cells. The fact
that high Rho and high Rac activity produces a mixture of possible coexisting
stable steady state cell shapes was independently predicted in a purely modeling
study by \cite{Holmes2016}. Related experiments by \cite{Sailem2014} on cell
shapes were also later modeled and explained in a follow-up paper by
\cite{Zmurchok2020}. These papers  demonstrate that relatively simple
stripped-down depictions of cellular signaling ``modules'' (such as Rac-Rho) can
account for important and unexpected observations at the level of the cell as a
whole.

The paper \cite{Mosier2019} explores how 3D collagen
microtracks and confinement affect cell migration. The authors find that
the degree of cell-extracellular matrix (ECM) interactions are key determinants
of speed, morphology, and cell-generate substrate strains during motility.

In recent times, a clear link has been established between GTPase activity and
mechanical tension experienced by cells. A pioneering experimental paper that
showed this 2-way feedback is \cite{Houk2012}.  The effects of forces on the
Rho family proteins, including the involvement of GEFs that activate Rac1, RhoA,
or Cdc42 are reviewed in \cite{Ohashi2017}. Some GEFs respond to cyclic stretch,
and others to tensile force or shear stress and substrate stiffness. Rac1 and
Cdc42 are activated by stretching of adhesion bonds \cite{Warner2019}. Rho is
mainly used in maintenance of focal adhesions, but it appears to play a
prominent role in cell--matrix interactions.

Strain and strain gradients affect cell orientation. Single cell experiments
with cyclic stretch are described in \cite{Chagnon2017}. The authors suggest
that the Rho-ROCK pathway that regulates myosin light-chain activity is
responsible for sensing and responding to strain gradients.

\LEK{Overall, it appears that the link between mechanical stimuli and GTPase
signaling is still young, providing ample opportunities for creative modeling.
So far, there is no consensus on what are the ``takeaways'' from the models so
far. Further, it would appear that this gap should be filled if we are to
successfully bridge between 1 cell and many, since the mechanics of cell
collisions, as well as cell collective migration, entail sensing of both chemical
and mechanical stimuli in the interacting cells. }

%
% Section 2: From single to collective cell behavior
%

\section{From single to collective cell behavior}

\begin{figure}[!h]
  \centering
 \includegraphics[width=0.9\textwidth]{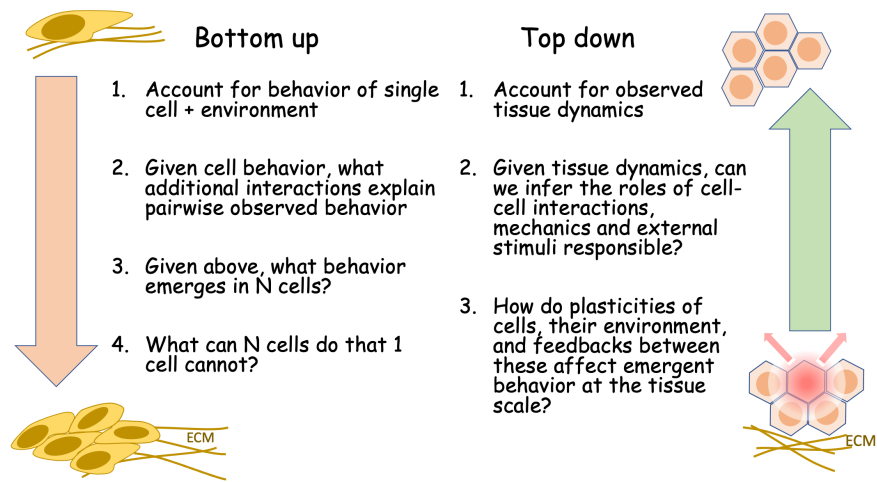}
\captionof{figure}{Modeling goals can be classified into broad categories that span
  levels of hierarchy. Some models attempt to span knowledge of single cell
  behavior plus interactions to predict emergent multicellular behavior
  (bottom up, left), whereas others start with observations of tissue dynamics
  and seek to infer underlying rules, feedbacks, and cell--cell (c--c)
  interactions that lead to those observations (top down, right).}
 \label{fig:BottomTop}
\end{figure}

While the meaning of ``collective behavior'' is intuitively clear, what is less
clear is how to specify the transition between a collection of agents, acting
individually, and the collective behavior of the group (right panel,
Fig~\ref{fig:BottomTop}).  To some extent, the same issue arises in
macroscopic models for swarming animals or interacting particles. As groups
grow and interactions between members increase, new distinct properties emerge
at the level of the group that were absent at lower levels of organization.
Characterizing, quantifying, and understanding such emergent properties remains
the single most interesting and elusive goal in bridging between single and
collective phenomena.

Unsurprisingly, physicists have grappled with similar questions in inanimate
systems. This goes 2 ways.  If we observe the dynamics of the group, can we
infer underlying interactions? This is a top-down approach; see
Fig~\ref{fig:BottomTop}. Conversely, given rules followed by individuals,
what can we predict about the structure and dynamics of the group? For example,
how do ``rules'' for cell--cell interactions such as coattraction, contact
inhibition of locomotion (CIL) affect cell collectives? In cell biology, we add
the question of how rules of behavior of cells are to be associated with
specific molecular/signaling pathways inside the cells.

From physics, we gain various concepts and techniques such as
order parameters, e.g.,\ average movement direction, normalized separation
distances, or migration persistence measures.  Mathematicians have also
contributed with useful methods, such as dynamical systems, ODEs, PDEs,
and bifurcation methods. More recently, the
use of topological data analysis (TDA) has also entered the mix, to help address
some of these questions. \LEK{The examples provided by
\cite{Topaz2015,Ulmer2019,Bhaskar2019,Mcguirl2020} show how TDA can be applied even more effectively
than the traditional physics-based order parameters to compare data and model
output. Such novel methods could help to begin addressing questions where model
hypotheses are to be compared with observed behavior. }

\subsection{Small cell groups}\label{subsec:SmallGroups}

A number of illuminating papers suggest that to understand the dynamics of
tissues, we should start by first understanding 1, 2, or a few interacting
cells in small groups \cite{George2017,dePascalis2017}. (See top right panel,
Fig~\ref{fig:BottomTop}.) Then, the behavior of single cells and the effect
of cell--cell interaction can be explored in detail. Collisions between cell
pairs or ``cell trains,'' contact inhibition of locomotion, and similar
responses fall into this category, as do ``cell swarms.''  The migration of
neural crest cells is 1 example of a loose cell swarm, as reviewed in
\cite{Schumacher2016}.  Adherent cells in tissues (and their cohesion in
epithelia) will be discussed in another section.

%%%%%%%%%%%%%%%%%%%%%%%%%%%%%%%%%%%%%%%%%%%%%%%%%%%%%
% Key questions that are asked in this area!       %%
%%%%%%%%%%%%%%%%%%%%%%%%%%%%%%%%%%%%%%%%%%%%%%%%%%%%%
A number of key questions posed by papers in this area include the following:

\begin{enumerate}[label=\textbf{(\arabic*)},ref=(\arabic*),leftmargin=*,labelindent=\parindent]
\item What mechanism, chemical and/or mechanical, can account for CIL in
interacting cell pairs \cite{Merchant2018}?
\item How do cells reconcile conflicting
cues \cite{Lin2015,Schumacher2016}?
\item Under what conditions would a cell reverse
its direction \cite{Lin2015}?

\end{enumerate}

The study of small numbers of interacting cells
provides a useful paradigm for connecting molecular mechanisms to group
behavior, an area that has only recently received the attention it deserves
\cite{dePascalis2017}.
\subsubsection{Bridging scales}

Phase-field methods have been used to model cell shape and motility in 2
spatial dimensions (2D) \cite{Camley2013}. The outline of a cell is represented
by a level curve of some
function $\phi(x,y,t)$. For example, in \cite{Kulawiak2016}, a minimal model for
Rac signaling was used to describe the shape and motility dynamics of a single
phase-field cell. The model is then extended to pairwise and cell--train
interactions in a narrow 2D strip. Interestingly, the authors had to assume an
intracellular signaling ``inhibitor'' that is activated at cell--cell contacts
to obtain appropriate behavior. This is one of  the earliest models spanning
the three levels of organization, from subcellular, to single and multiple
cells. The model exploits a reasonably simple level of detail at each stage to
reproduce several behaviors, such as reversals,  cells walking-past one
another, cell sticking, and forming a ``cell--train.'' \LEK{We feel that this
paper provides good prototype to be emulated by community members: a clear and
well-studied intracellular system, linked to evolving cell shape, with feedback
from cell--cell interactions. The paper leads to natural follow-up questions,
amenable to experimental investigation: what cell components play the role of
the putative ``inhibitor''?}

Models for neural crest cell (NCC) swarms in vivo and in vitro were proposed in
\cite{mclennan2012,mclennan2015,Schumacher2016}, focusing on distinct behaviors
of leader and follower cells. (See ``Links with experiments'' for a summary.)
For the same NCC system, the paper \cite{Merchant2018} developed a multiscale
computational model linking signaling to cell group migration. A Rac-Rho GTPase
circuit was represented by differential equations (ODEs) at nodes on the
perimeter of a deforming polygonal cell. Co-attraction was represented as a
non local tendency of cells to cluster, short range contact inhibition by
up-regulating Rho at nodes in contact with other cells.  Stochastic noise in Rac
activity generated characteristic run lengths and reorientation seen in typical
NCCs.

In \cite{Merchant2018}, it was shown that these underlying biochemical systems
could account for CIL. This study can be compared with the GTPase and inhibitor
system described above, \cite{Camley2014}. Coherent migration of a group of $n$
cells, each of diameter $d$, was possible along a confinement corridor of width
on the order of $d\sqrt{n}$. This paper,  \cite{Merchant2018}, has a wider
corridor and larger numbers of cells than \cite{Kulawiak2016}, but confinement
serves a similar function of enhancing directed cell movement.  In real embryos,
such ``corridors'' could be defined by permissive and inhibitory regions for NCC
migration, as shown experimentally in developing embryos \cite{Kulesa2010}. As a
comparison to the models of \cite{Merchant2018,Kulawiak2016}, in
\cite{Woods2014} a ``rule-based'' approach was applied to the same problem of
NCC group migration, without bridging to the biochemical circuit.

In somewhat related work, \cite{Zhao2020}, 2D cells are described by
viscoelastic triangular finite elements. Cells can adhere to
one another using linear springs. At each cell vertex, intracellular
signalling controls cell protrusion and cell--matrix adhesion.
The signaling networks consists of Rac
\LEK{and several other components, including the downstream kinase PAK, the
focal adhesion protein Paxillin, the cell adhesion molecule cadherin, and
Merlin, a protein implicated in contact inhibition of cells in a feedback loop
with Rac1 and PAK.} Key findings of this study were (1) cells elongate on
stiffer substrates; and (2) stiffer substrates allow for better
directional guidance and collective cell migration. Interestingly, the authors
state that the principal role of intracellular signaling is to coordinate cell
movement direction over distances of $\approx 500\mu m$.

\subsubsection{Levels of detail}

There are many ``simple'' SPP models in the literature
on collective cell behavior, reviewed, for example, in
\cite{Mehes2014,Camley2017b}. It is relatively straightforward to create rules
that result in collective behaviors, but harder to do so when more details are
included.

In 1 SPP cell model in a 1D spatial domain,  \cite{George2017}, CIL and
force-induced cell repolarization result in properties of collective cell
behavior. Rules governing pairwise interactions and probabilities for CIL
are derived from experiments by \cite{desai2013} described in the next section.
It turns out that there is an optimal number of cells in a group for migration
persistence. This paper joins \cite{desai2013,Lin2012,Lin2015} and others in
proposing and using simplified 1D geometry to examine the basics of cell--cell
interactions. \LEK{The synergy between papers \cite{George2017,desai2013} is very helpful, and points to the need for more paired experiment-model studies.}

SPP models are easily extended to higher dimensions, and/or
more realistic pairwise forces. For example, \cite{Carrillo2018} considers a cell
swarming model in 2D that is closely related to  spherical cell models in
\cite{Drasdo2005,Hoehme2010}. While \cite{Carrillo2018} neglects cell--cell
friction, the paper couples
the cells' compressibility, which plays the role of a local force of repulsion,
to adhesion due to long protrusions such as filopodia that leads to nonlocality.

\LEK{The highlight of \cite{Carrillo2018} is the inclusion of these non local
cell--cell attraction forces. In this way, the paper makes a seminal link
between extended cell--cell interactions and a body of theoretical work on
non local interactions in swarms or physical particles.  The same paper also
features the
concept of H-stability (preservation of bounded density as the number of
particles increases). This idea is used to ascertain correct cell
spacing in swarming models, facilitating  links to experimental observations.
This paper hence imports a panoply of techniques, methods of analysis, and
results, from theoretical work on non local models dating back to the 1990s
\cite{Mogilner1999}, and inspires new ways of interpreting experiments.}

Parallels with macroscopic swarms and flocks are useful, so long as we remember
that these operate in vastly different friction regimes. (Cells experience an
overdamped regime, where inertial forces are negligible.)
In yet another example of a cell swarming model, the paper \cite{Volkening2015} accounts
for the development of stripes in zebrafish pigmentation patterns. The model
comprises 2 cell types, with cell differentiation and death, and  successfully
reproduces a wide array of observed wild-type and mutant patterns.

In a different approach at a lower level of detail, \cite{cai2016} describes a
group of border cells in Drosophila as a single sphere to determine the role of
cluster size in chemotactic migration speed. The authors assume a spherical
cluster moving by Stokes' law in a viscous environment. The migration force is
proportional to cell surface receptor occupancy in an exponential gradient of
attractant. The results show good fits to both normal cells and cells that have
a deficiency of receptors.
The same biological problem is treated in \cite{Stonko2015}, using a force-based
approach for individual spherical cells (rather than cluster as a whole as in \cite{cai2016})
in 2D, followed by 3D (for small clusters of up to 8 cells). The computation
incorporates cellular adhesion and repulsion, with some stochastic effects.

At the next level of detail, the shape of interacting cells is explicitly
represented.  The papers \cite{Shao2010,Camley2014} apply
phase-field methods to model cell shapes in 2D with intracellular signaling.
Interestingly, the signaling system is itself ``minimal:'' The authors assume
basic ``wave-pinning'' GTPase dynamics, as in \cite{Mori2008}, with additional
stochastic elements. The levels of internal GTPase give rise to either
protruding or retracting forces on the cell edge.  Then, \cite{Camley2014}
models the rotational cell motion when 2 cells interact (compare with
\cite{Rens2019} who showed a similar behavior in CPM cells). For coordinated
movement, the directions of these gradients need to be linked.  A key finding of
\cite{Camley2014} was that the emergence of rotational motion is strongly
dependent on the type of cell--cell coordination mechanism (which includes the
previous ``inhibitor'' field and other phenomenological cell polarization
mechanisms).

We see gradual ascent in the numbers of cells considered and level of detail. We
now find groups addressing the question ``What can cells do well together that
they do poorly on their own?'' For example, collective chemotaxis was shown to
be better in groups than in single cells \cite{Camley2017,Camley2018,Merchant2020}.
(See also the related work by \cite{Lober2015} for larger cell groups, reviewed
in the next section.)

\LEK{Overall, our sampling of the literature reveals a paucity of models at the
small-cell group level, relative to subcellular and tissue-scale modeling. We
believe that papers such as \cite{Carrillo2018} open opportunities for
swarm-centered modelers to make a contribution to the cellular biology realm, by
translating numerous results of physics and mathematics, suitably reinterpreted,
to cells. We also contend that understanding the ways that cell interact in
small groups is an important and relatively tractable step to formulating more
informed tissue-scale models.}

\subsubsection{Links with experiments}

Pairwise interactions of MTLn3 rat breast adenocarcinoma cells inside
microfluidic channels with chemotactic cues were quantified in \cite{Lin2015}.
These experiments provide important information for the spatio temporal
evolution of GTPase activity during cell collisions. When cells collided in
\cite{Lin2015}, they either stalled, adhered and moved together, or exhibited
CIL. \Andreas{More recently, \cite{Jain2020} studied how mechanics, and cell
polarity drive collective cell migration in larger groups of cells}. More
experiments of this type would be very important in laying the groundwork for
understanding pairwise cell interactions and in directly informing the
development of models.  Such models could then be used to interpret and deepen
experimental observations.

In a similar vein is \cite{desai2013}, a study of CIL experiments on
micropatterned surfaces, where data were collected for the probability of cell repolarization
after contact with another cell. The authors also explored persistence of
cell trains, bridging from pairs to larger groups. An agent-based model was then
used to predict cell--train outcomes based on the observed repolarization
probabilities.

A model that replicates the experimental setup of \cite{Lo2000} appears in
\cite{Schluter2012}.  Flat cylindrical model cells are assumed to align ECM
fibers.  Primarily a single-cell model, it is also extended to 2 interacting
cells where, interestingly,  leader-follower behavior is seen: The leader
creates a ``track'' of aligned ECM fibers that the second cell can follow.

A number of works shed light on the molecular basis of cell--cell interactions.
\LEK{For instance, Wnt signaling is a ubiquitous pathway that regulates
 polarity, migration, and many other cellular processes. Wnt signaling at
 cell--cell interfaces that locally upregulates Rho is studied in
\cite{Carmona2008}.  Ephrins form cell--cell ligand--receptor bonds that funnel
signals to GTPases to control cell repulsion or attraction.} Involvement of
Ephrin receptors in CIL \cite{Astin2010}, effect of tension on CIL
\cite{Roycroft2015,Davis2015}, and on the activation of RhoA \cite{Lessey2012}
have been similarly studied.

Experiments on migration of primordial progenitor cells in the developing
zebrafish \cite{Paksa2016} established the importance of guidance cues such as repulsion due to ephrins. Physical barriers also guide
cell migration and organ placement. Actin was visualized for cells colliding
with barriers expressing ephrin receptors. A particle-based model with an
attraction--repulsion Lennard-Jones potential was then used to test the effects
of reflecting and nonreflecting boundaries. \LEK{(The Lennard-Jones potential is
composed of power functions of the form $r^{-12}$ and $r^{-6}$ for local
repulsion and long-range attraction.)}

NCCs chemotax toward some cell types in a chase-and-run behavior
\cite{Theveneau2013}. \LEK{For example, NCCs chase the placode cells, namely cells that are destined to form feathers or teeth or hair follicles.}
Before contact, each cell has high Rac at its front. Upon
contact, \LEK{the cell-adhesion protein} N-cadherin, at the cell--cell interface, inhibits Rac in both cells
leading to separation and escape of the placodes.  Once the cells separate,
chemotaxis is reestablished.

The group of Danuser \cite{Ng2015} used high-resolution traction-force
microscopy (TFM) to measure traction forces for a single cell and small groups
of 2 to 6 epithelial cells. They then used a thin-plate finite element model
(assuming homogeneous elasticity) to reconstruct forces along the interfaces
between cells in a cluster. It was shown that the forces were correlated with
E-cadherin localization. The authors reduce the cell group to a network
representation, with vectors connecting cell centroids.

\Andreas{A recent paper explores the relationship between initial tissue size and the dynamics of tissue growth \cite{Heinrich2020}.}

Experiments on the migration of chick neural crest cells in vivo and in vitro
are described by several groups, including that of Kulesa. His experiments
have been directly linked to models previously mentioned
\cite{mclennan2012,mclennan2015,Schumacher2016}. The group has investigated how
cells reconcile mixed chemotactic signals, as in breast cancer cells in \cite{Lin2015}, and how
information is shared between cells. The differences between
leaders and followers is of recent interest in the group.

\LEK{While papers such as \cite{Capuana2020,George2017} argue eloquently for the
need to carry out small-group cell experiments, we find this, too, to be a
missed opportunity. Possibly, the in vitro work that this entails is viewed as
less compelling than in vivo experiments, and so, less persuasive from the
grantsmanship perspective. Hence, there are still relatively few experiments to
map out the links between cell signaling and cell--cell collisions, between
events inside the cell and the small-scale swarms that form. We recommend this
area as a promising one for experiment-model synergy.}

\subsection{A related example: Spacing of nuclei}

While not directly related to cell migration, we nevertheless decided to briefly
highlight the following example as it illustrates close synergy between model
and experiment using modern methods.  In \cite{Manhart2018}, models complement
experiments to investigate positioning of nuclei in Drosophila larvae muscle
cells. The authors used data for many hundreds of nuclei and proposed a
diverse set of particle-based models. Models were formulated specifically to
test distinct sets of hypotheses for how nuclei interact with microtubules and
molecular motors. Interestingly, machine learning was used for computational
filtering, comparing large numbers of simulation outcomes to simple summary
statistics such as nuclei positions. The filters are applied in stages, each
stage rejecting inappropriate or inaccurate models that are inconsistent with
the data.

Ideally, this process requires high-throughput screening, mandating both
clear-cut summary statistics, and a large dataset.  As noted by
\cite{Manhart2018}, the availability of data imposes a limitation on possible
model complexity that could be adequately fit.  Eventually, the best-fit model
predicts that nuclei push against one another and against cell boundaries by
sending out growing microtubules.

This example suggests a number of important questions for future
model-experiment investigations. (1)  How can one ascertain the optimal level of
model complexity in the first place? (2) How do we extract clear identifiable,
meaningful summary statistics from a large dataset?
(positions of hundreds of cells) (3) What are optimal
methods for massive filtering to obtain the best model? \LEK{This example also
illustrates a creative use of machine learning as a tool in understanding and
filtering many possible mechanistic explanations. Clearly, this approach goes
beyond the common machine learning application as a ``black box'' to sort or
classify or learn rote features of cells or images.}

%
% Section 2.3: Collective cell behavior and tissue dynamics
%
\subsection{Collective cell behavior and tissue dynamics}\label{subsec:Tissuebehavior}

\LEK{Even before details of cell--cell signaling were known, theorists were
formulating cell-based computational models for tissue dynamics. Garret Odell and
George Oster \cite{Odell1981} considered cell--cell pulling
forces, coupled to a bistable stretch-sensing signaling module. They showed that
this mechanism could account for local contractions that folds a tissue in a
developmental process such as gastrulation. Their seminal idea still resonates
today \cite{Zmurchok2018}, now that signaling components are familiar to us.  }

Reviews of the recent biological literature on collective cell migration include
\cite{Scarpa2016,Friedl2009,Haeger2015}. (See also brief highlights in
Table~\ref{Table:ExperimentSummary}.) A thought-provoking review of the biology,
 \cite{dePascalis2017}, focuses on the role of adhesion mechanics, bridging
between force regulation in single cells and in collective cell migration. The
paper gives detailed summary of the underlying molecular players, synthesizing a
vast literature on the subject. The mechanical and mechanotransduction players
at the molecular level are then linked to consequent cell-level properties.
The paper proposes a rational program   of experiments
from 1 to 2 to many cells (see Fig~4 in \cite{dePascalis2017})
to bridge the scales from single cells to tissues.

\begin{table}[!h]
  \centering\tiny
  \begin{tabularx}{\textwidth}{@{} X X t @{}}
    \toprule
   \bf Main-Target & \bf Main-Finding & \bf Ref. \\
    \midrule
     Rac opto-activation  & Cell migration by photo-activation (HeLa cells) &\cite{wu2009} \\

    Cdc42 opto-activation  & Cdc42 activates
     Rac at the front, and Rho at the back of the cell (Immune cells) &   \cite{ONeill2016}  \\

     Rho opto-activation  & Rho
     at cell's rear can control directional migration.
     Rho activity regulates switch between amoeboid and mesenchymal migration
     (macrophages) & \cite{ONeill2018}  \\

     Rho-ROCK
     pathway & This
     pathway senses and responds to strain gradients; cyclic stretching (single cells)&  \cite{Chagnon2017}  \\

     Chemotactic response & If gradient switches too rapidly, cells get
     stuck (Dictyostelium) &  \cite{Meier2011} \\

      Cell polarization &  Comparison of three RDEs, parameter fitting to data &  \cite{Lockley2015}  \\

      Merlin  &   Merlin is a negative regulator
     of Rac and may also be regulated by the Rho pathway &\cite{Bretscher2002}    \\

      Merlin & Key role in cell polarity, and leadership. Spatio-temporal data (cell monolayers)&  \cite{Das2015}  \\

     Signaling at cell-cell interfaces.  & Non-canonical Wnt signaling at cell-cell contacts causes an upregulation of Rho & \cite{Carmona2008}   \\

     Forces and GTPase activity & Cell-cell forces affect Rho activation &  \cite{Lessey2012}  \\

     Topographic cues & Single cells
     have different organizations on grooved vs flat substrata (fibroblasts)&   \cite{Londono2014}  \\

     NCCs and placode cells & Intricate interplay between chemotaxis, integrins and their effect on internal GTPase activity &  \cite{Theveneau2013}  \\

       Ephrins & Ephrins in repolarization of cells (zebrafish development) & \cite{Paksa2016}  \\

     CIL and Ephrin receptors & Ephrin receptors affect CIL &  \cite{Astin2010}   \\

     CIL  & Biphasic relationship between probability of CIL and
     collective migration &  \cite{desai2013}  \\

    CIL & Tension built up during CIL &   \cite{Roycroft2015, Davis2015} \\

      CIL &  Spatio-temporal GTPase patterns during cell-cell collisions and CIL events (microfluidic channels) & \cite{Lin2012,Lin2015}  \\

      Collective strand formation & Interesting difference
     between polarization of leader and followers &   \cite{Reffay2014}  \\

    Small groups of epithelial cells & Forces are
     correlated with E-cadherin localization &   \cite{Ng2015}   \\

      Effect of forces on Rho and Rac GEFs &  Some GEFs respond to cyclic stretch, others to
     tensile force or shear stress and substrate stiffness & \cite{Ohashi2017}  \\

     Mechanical GTPase activation  & Rac1, Cdc42 activated by stretching adhesion bonds. Rho  maintains focal adhesions &   \cite{Warner2019}  \\

     Epithelia &
     ``Push-pull'' or ``caterpillar'' collective motion in narrow grooves, more complicated movement in wide channels &   \cite{Vedula2012}  \\

      Epithelia & Sustained oscillations in epithelial sheets & \cite{Peyret2019}  \\

 Epithelia  & Cells crawl in the direction
    of maximal principal stress. Leaders impose
    mechanical cues on followers &      \cite{Zaritsky2015}   \\

  GTPases and GEFs  & Coordinated waves of motion, cells at front
    react first (scratch-wound assay in human bronchial
    epithelial cells) &    \cite{Zaritsky2017}  \\
    \bottomrule
  \end{tabularx}
  \caption{Experimental summary: CIL, contact inhibition of locomotion;
   GEF, guanine nucleotide exchange factor; NCCs, neural crest cells; RDEs,
   reaction diffusion equations.
  }\label{Table:ExperimentSummary}
\end{table}

%%%%%%%%%%%%%%%%%%%%%%%%%%%%%%%%%%%%%%%%%%%%%%%%%%%%%
% Key questions that are asked in this area!       %%
%%%%%%%%%%%%%%%%%%%%%%%%%%%%%%%%%%%%%%%%%%%%%%%%%%%%%
Several key questions appear in papers at this level.

\begin{enumerate}[label=\textbf{(Q\arabic*)},ref=(Q\arabic*),leftmargin=*,labelindent=\parindent]
\item How do internal dynamics of cells influence the emergence of collective behavior? How do cells share
 and transfer information between one another? How and to what extent
are mechanical forces transmitted over distance \cite{Shi2018}?
\item What is the role of the leading front in guiding collective migration? \cite{Vedula2012}
\item Do actin cables or ECM fibers transmit long-range stresses
\cite{Reffay2014,Spatarelu2019}?  Are such mechanisms analogous to the non local
sensing mechanisms that play a dominant role in models of animal swarms?
\item What cellular and tissue-based mechanisms (long- and short-range cues, barriers,
cell--cell adhesion, etc.) are essential for proper formation of organs during
development \cite{kasemeier2015trkb,Paksa2016}?
\end{enumerate}

Parallels between the polarization and directed migration of single cells and of
a cell collective is highlighted in many recent works. See, for example, the recent perspective
paper, \cite{Capuana2020}, demonstrating parallels in the GTPase distribution,
chemotaxis, and mechanosensing between the single cell level and tissue level.

As shown in Fig~\ref{fig:CellSortingAndMono}, many computational platforms
are currently used to simulate multicellular migration. These are reviewed in
\cite{vanLiedekerke2015,Camley2017b,vanLiedekerke2018,Yang2019,Alert2019}.
Rigid or deformable spheres or ellipsoids (in 3D or 2D,
Fig~\ref{fig:CellSortingAndMono}A1 and~\ref{fig:CellSortingAndMono}B1, respectively), polygonal
``vertex-based,'' or CPM cell representations are common. Software platforms such
as CHASTE (Oxford University, \cite{Fletcher2013,Osborne2015, Osborne2017}, and
Fig~\ref{fig:CellSortingAndMono}A2), CompuCell3D (U Indiana, \cite{Swat2012},
and Fig~\ref{fig:CellSortingAndMono}A3 and~\ref{fig:CellSortingAndMono}B2), or Morpheus (TU Dresden,
\cite{Starruss2014}) have made it increasingly easier to simulate complex tissue
dynamics without having to reinvent computational algorithms and graphics.

\LEK{In our humble opinion, there is a vital need for support and sharing of
standardized open-source software packages, with suitable plug-ins donated by
groups utilizing those resources. First, such tools would save time,
person-power, expertise, and expense of individual ``kludges.'' It would reduce
duplication of effort --- how many of us want to reinvent our own finite-element
computations? Even more compellingly, such standardization can bring about much
easier communication and appraisal of published models, and in-depth scrutiny of
exactly what those models include. Having learned the basic steps of Morpheus,
one of us (LEK) has become an enthusiastic user. See, for example,
\cite{Mulberry2020}, where figures are linked directly to executable Morpheus
\texttt{.xml} model files that produced them. }

\subsubsection{Bridging scales}

Here we concentrate on bridging between models for biochemical signaling and
multicellular behavior.  In the next section, we will focus on bridges between
1 or a few cells to larger populations and from simple to more detailed
geometry (points, spheres, and cell shapes) and dimensionality (1D, 2D, and 3D).

The modeling paper by \cite{Koride2018} is a vertex-based computation of
epithelial dynamics, where force balance of cell vertices includes an active
contractile force on the cell perimeter due to myosin. ODEs track Rho activation
by cell perimeter stretching and downstream myosin activation (by
phosphorylation) at a given vertex. Vertex motion is a result of force
balance. Friction and passive forces are derived from energy penalties for area
constraint and from the energy of adhesion to neighbor edges. This aspect of the paper
resembles the energy-based approach in CPM computations, e.g., in
\cite{Zmurchok2018}. The authors quantified the effects of cell density on
correlation lengths, and regimes of behavior such as streaming, contractile
waves, group movement, rotation in a circular domain, presence of vortices, and
non uniformity of myosin distribution.

Some similarities are shared by \cite{Koride2018,Zmurchok2018}. Both are
tissue-based computations, with ODEs for GTPase biochemistry, an assumption that
GTPase leads to cell contraction or expansion, and feedback from cell tension
back to the activation of the GTPase.  Both papers find fluctuations in cell
shape or size from the full feedback between signaling and contractility (Fig~2C
in \cite{Koride2018}, and the comparable Fig~3B in \cite{Zmurchok2018}). The
signaling biochemistry is assigned to vertices in \cite{Koride2018} versus the
cell interior in\cite{Zmurchok2018}. In \cite{Zmurchok2018}, a single cell is
first linked to a 1D chain of cells, and then to 2D CPM cells, where waves of
cell contraction are observed. The levels of detail in these 2 papers are
comparable, though \cite{Koride2018} also includes predictions for myosin, unlike
\cite{Zmurchok2018}.  See also \cite{Bui2019} for a similar approach restricted
to 1D cell chains.

In the same general class, the work of \cite{Durney2018} links cell contraction
(in a vertex-based hexagonal cell with 6 spokes) to biochemical details
for a circuit of proteins known to regulate myosin contractility in Drosophila
dorsal closure. This model is more detailed on the specific known interactions
of protease-activated receptor (PAR) proteins \LEK{that form a negative feedback
loop with actomyosin} (Baz, Par-6, aPKC), including a set of 9 ODEs for
components linked
to myosin dynamics along cell edges and spokes. The authors account for several
phases in the developmental process, including oscillations observed in the
tissue (using time delays) and the eventual contraction of the tissue.
By way of comparison, \cite{Jamali2010} has a greater level of detail of the
cell shape, including more viscoelastic spokes and edges, but essentially no biochemical
detail.  An example of 3D vertex-based simulations with
cell rearrangement and out-of-plane bending can be found in \cite{Osterfield2013}.

\subsubsection{Levels of detail}

The gene regulatory network involved in Drosophila ventral furrow formation is
modeled at a fine level of molecular detail in  \cite{Aracena2006} using a
Boolean modeling approach. The model encapsulates interactions of  transcription
factors to explain 3 phenotypic attractor states, and identify missing and
alternative pathways. Spatial distribution of cells and cell shape dynamics is
not considered.

At the opposite extreme, internal cell signaling is omitted in favor of
positions and spatiotemporal dynamics of cells. SPP ``agent-based'' models with
forces of attraction, adhesion, and repulsion include \cite{Carrillo2018},
 as reviewed in a previous section. There, nonlocal forces are considered in groups
of hundreds of cells.

Modified agent-based models in \cite{Tarle2017} can be compared to the cell-pair
study in \cite{Camley2014}.  Additional assumptions are included in
\cite{Tarle2017}, notably Vicsek-type alignment, cell membrane curvature,
elasticity, actomyosin cable
force, and a tendency to move in the outwards normal direction. The authors show
the dependence of tissue wound-healing velocity on the width of a confining
stripe. On a similar topic, CPM computations are used in \cite{Albert2016} to
investigate the effect of adhesive micro patterns on the motion of single cells
and collective cell migration.
Similarly, modified Viscek-type models are considered by \cite{George2017}
endowing each cell with a polarization direction. Various types of cell--cell
polarity coordination mechanisms are considered.

Yet another modification of particle-based models is \cite{Schnyder2017}, where
a cell is represented by a spring linking 2 discs, to depict the ``cell body''
and the ``pseudopod.'' The model demonstrates various outcomes of binary
collisions in a 2D domain, as well as order--disorder
transitions and velocity waves in large 2D cell groups with and without CIL.
This paper explicitly bridges from 1 to many cells.

At the next level of complexity, we find representations of cells as spheroids
or ellipsoids (Fig~\ref{fig:CellSortingAndMono}A1,
\cite{Drasdo1995,Palsson2000,Drasdo2005,Palsson2008}). Drasdo computed a 2D
growing monolayer \cite{Drasdo1995}, and later extended it to 3D
\cite{Drasdo2005}. Palsson built 3D simulations of deformable off-lattice
ellipsoids, with cell division, chemotaxis, cell--cell adhesion, and
volume exclusion. The platform has been used to model the dynamics of
Dictyostelium aggregation and cyclic AMP (cAMP) signaling \cite{Palsson2000}, as
well as the posterior lateral line primordium, a cluster of about 100 migrating
cells in zebrafish embryos that generate sensory structures on the surface of the fish \cite{Knutsdottir2017}.

Other recent papers add variations on these themes. For example,
\cite{Frascoli2013} provides a computational model for 3D off-lattice spheroid
cells, but with greater level of detail for cell--cell and cell--matrix adhesion. The model is
coarse grained from the biological levels of $10^6$ down to 100 sites per
cell. After quantifying the stable distance between cell pairs (for a given set
of forces, intrinsic cell properties, and adhesion sites), the authors go on to
show the behavior of a larger group of cells.

Recently, these approaches have been extended to more finely resolve cell shapes
in 3D \cite{Liedekerke2020}. Here, each cell is a triangulated 3D object, composed of discrete viscoelastic elements. Simulations track up to a thousand
liver cells. While computationally very expensive, these approaches can provide
valuable insights.  For example, it is shown that highly deformable cells move more easily to
heal a lesion than stiffer or more rigid cells \cite{Liedekerke2020}.
They are also useful for specific systems where it is crucial to capture details
accurately, as, for example, in drug design in silico for liver disease.

In \cite{Lin2018, Lin2019}, a confluent epithelium in confinement is modeled by
vertex-based polygonal cells, and the type of collective migration (directed
flow, vortex chain, or turbulent) is related to a dimensionless quantity (``cell
motility number'') that combines cell motility, cell density, and size of the
confinement pattern.

In the phase-field category, the paper by \cite{Lober2015} scales up from small
numbers of cells in \cite{Camley2014} to larger populations.  Cell collisions
are modeled by an energetic penalty for overlapping phase fields, and cell
adhesion is described by a reaction--diffusion equation in \cite{Lober2015}.
Their model cells exhibit bistable shapes, (either symmetric or
keratocyte-like),
and both elastic and adherent collisions are predicted. Strong adhesion results
in bands of dense closely packed cells that move as a collective. The authors
compare their work with that of \cite{Doxzen2013} (see Links with experiments)
where a CPM model is used to describe large confluent cell rotations in a
confined region. As noted by \cite{Lober2015}, the level of detail in their
phase-field description is better suited for non-confluent cell models, but
likely too detailed for confluent layers of cells where individual cell details
average out. In \cite{Winkler2019}, similar methods are extended to a 3D phase
field computation for a single cell interacting with curved or grooved surface.
This sequence illustrates the  trade-off between geometric complexity (1D versus
2D versus 3D) and collective complexity (1 cell versus few versus many).

\Andreas{Overall, it emerges from many papers that the vertex-based models are
good at describing epithelia, where fragmentation or EMT is absent or not
important. Vertex-based models are not well suited to track cell death or
fragmenting clusters, since edges and nodes are shared by neighboring cells.
Breakage of a cell requires that shared edges and nodes be duplicated or
reassigned an inconvenience. To track tissue fragmentation or loss of cells,
center-based models or CPM platforms have an advantage, since these represent
each cell by an individual geometric object or by a set of pixels.}

Continuum models are also commonly used at the level of tissue dynamics. Here,
the cell identity is omitted altogether, in favor of cell densities, local
flows, and material properties. Analogies are made with fluids, viscoelastic
material, elastic sheets, or foams. Advantages include the ability to harness
traditional methods of physics, mathematics, and fluid or engineering
computational software.  Examples of this approach are numerous, and we mention
only a few.

One example of this class, \cite{hannezo2014}, is a 3D physical description
of an epithelial sheet. The authors show bistability in  cell aspect ratio,
bending and buckling instabilities of epithelia, and transitions that lead to
formation of epithelial tubes and spheres. Experimentally testable scaling laws
are given for such morphogenetic transitions. While there is no subcellular
molecular detail, the connection between local cell shape and epithelial
behavior is an important contribution.

In \cite{Banerjee2019}, the tissue is modeled as an active continuous medium,
with a Maxwell viscoelastic constitutive law. A reaction--diffusion-transport
equation accounts for actomyosin, whose contractility is coupled to cell motion
and polarity. The authors explore predicted motion of tissue in confined
regions, closure of wound gaps, and the relationship between traction forces and
sizes of cohesive cell clusters. Mechanical waves, as well as tissue
rigidity cycles, are observed, where fast fluidization is followed by a gradual
period of stiffening.

The continuum model in \cite{Arciero2011} treats a tissue as a compressible
fluid and includes both cell division and death in a type of Stefan
free boundary problem. The authors show that behavior of the tissue in wound
closure and in colony expansion depends on 3 parameters: 2 physical constants
and the proliferation rate. They argue that all these can be estimated from limited
experimental data, with examples calculated for \LEK{IEC-6, a rat cancer cell
line and MDCK cells, a canine kidney cell line}. As the authors point out,
continuum models are appropriate, provided the size of
the tissue or the characteristic length of the wound is sufficiently large
relative to the size of individual cells.

In some cases, an approach that combines features of both continuum and discrete
cell identity is implemented. One example is \cite{Escribano2018}. Here, a 1D
cell monolayer consists of a long contractile element with myosin creating a
strain rate resulting in length changes. This element is flanked by cells in
front and rear. Binding and unbinding of adhesion sites are also included. The
model tissue exhibits durotaxis,  movement directed up a stiffness gradient, under
appropriate conditions on the myosin and adhesion parameters. The authors use
this model to conclude that a monolayer is more effective at durotaxis than
single cells.

\LEK{Other examples of hybrid treatments of particle-based and continuum
approaches include \cite{Degond2018, Degond2020}. These papers tackle the
important question of how to derive appropriate continuum models from underlying
SPP models. This kind of work and, in particular, expository
papers that summarize the conclusions in ways that biological modelers can
understand, could help link work in the literature that is currently
underappreciated or not understood. }

A recent work, \cite{Yang2020}, describes fingering at the front of an epithelial
sheet using several of the above approaches. Cells are represented by pairs of
points, as in \cite{Schnyder2017}, and also by Voronoi polygons for computations
corresponding to experiments in \cite{Reffay2014}. The authors develop an
active fluid model for the epithelium. Using these combined approaches, they
demonstrate that stable fingering of the tissue edge requires leader cells. They
characterize the wavenumber of the fingering instabilities (distance between
stable fingers) using stability analysis of the continuum PDE model.

\LEK{In summary, while the literature on tissue-scale collective migration is
rapidly growing, it is still in stages of infancy as far as coherence,
coordination, and clear direction are concerned. We see many exploratory steps
in many individualistic directions. The list of core questions and the
discovery of unifying principles is beyond the horizon, providing ample
challenges for the community. We can draw an analogy between the current state
of the art and the behavior of a growing but uncoordinated population of cells.
There is no emerging unified front.  Each is exploring independently and
occasionally aggregating with a few others, but the global organizing principles
are yet to emerge.}

\subsubsection{Links with experiments}

The links between molecular players such as Rho family GTPases and epithelial
morphogenesis have been known for some time. (Experimental literature reviewed in
\cite{vanAelst2002}.) The role of Rho GTPases in the leader-follower identities
and in front--rear tissue polarity is reviewed in \cite{Zegers2014}.

Over the years, investigators have been asking how  external constraints,
geometry, strain fields, gradients, and other factors affect collective
behavior. How does collective cell migration emerge from transfer of mechanical
information between cells? For a cell to be a ``leader,'' should it be more
sensitive to stimulation than other cells? Should it have elevated or more
responsive GTPase activity, for example?

These and similar questions are posed in the experimental work of \cite{Zaritsky2015}. Here, the authors investigate
the roles of the GTPases RhoA and RhoC and their GEFs in collective cell migration. It is found that cells tend to crawl in the direction of maximal
principal stress, a process called plithotaxis. In their scratch-wound assay of
MDCK and human bronchial epithelial cells, wounding leads to coordinated waves
of motion, with cells at the front edge reacting first, followed by those
successively further back. The authors speculate that mechanical cues are
induced by leader cells on followers behind then by normal strain, and alongside them by
shear strain to coordinate motion.

In \cite{Zaritsky2015}, the relationship of tissue speed to distance from the
leading edge is quantified using particle image velocimetry. The authors employ
shRNA to knock down GTPase Cdc42 or Rac1 and many of their GEFs (screening some
81 GEFs in total). They find that the speed and directionality of the cells drop
everywhere. When RhoA is depleted, there is a reduction in the spatial gradient
of cell speed.
They suggest that the molecular mechanism may include signaling
downstream of cadherin, as well as Merlin-Rac1 signaling.

The experiments in \cite{Das2015} elucidate the effect of pulling stress on Rac
and Rho GTPases.  The authors investigated which properties of underlying
molecular machinery allow for coupling between mechanical forces and correlated
cell motion.
The correlation length scale of collective force transmission was determined
experimentally in \cite{Vishwakarma2018}. Observed behavior was then modeled
using a thin elastic sheet of height $h$ with some elasticity and isotropic
contraction stress. The authors investigate the emergence of leader cells and
found a typical length scale  of about 170 $\mu$m.

\LEK{A number of experimental studies have explored how epithelia behave in grooves \cite{Vedula2012,Londono2014},  various topographic surfaces, or confined settings \cite{Jain2020}, including arrays of posts \cite{Wong2014}, arrays of grooves \cite{Park2019}, or adhesive \cite{Doxzen2013} surfaces. }
In \cite{Vedula2012}, the width of grooves and adhesive strips is varied to investigate how these affect
 collective migration of an epithelial sheet of MDCK cells. The authors map out the force and velocity fields in each case.
Their narrowest grooves are 20 $\mu$m wide, so cells are in a single file, and
the authors observe ``push-pull'' or ``caterpillar'' type motion, resembling the
relaxation--contraction cycles in the model by \cite{Zmurchok2018}.  Larger
tissue widths exhibit vortices of cell motion. The authors deduce that the
constraints of the geometry influence cell rearrangement, as well as junctional
forces between cells. (Compare with \cite{Peyret2019}, who observed sustained
oscillations in epithelial sheets.)

Micropillar arrays form the playing field in \cite{Wong2014} to observe
collective migration and EMT in mammary epithelia and breast cancer cell lines. Dispersal of single, highly motile
mesenchymal cells from a spreading  epithelial front was shown to agree
quantitatively with a minimal physical model of binary mixture solidification.
While such a model lacks cell detail, it has several advantages, including
simplicity, basic summary statistics for comparison with experiments, and
availability of an analytic solution. Methods from physics can be used to inform
the link between the material properties and the overall macroscopic behavior.

Both single fibroblasts and epithelial monolayers were studied in
\cite{Londono2014}, showing that  cells tend to be more organized on grooved
versus flat substrata.  Mechanical exclusion interactions, rather than strength
of junctions between cells, affect the distance over which the topographic
guidance signal propagates between cells. The same paper proposes a CPM
computational model to describe the observations. In the CPM model, following
the the style of \cite{Kabla2012}, each cell is assigned a phenomenological
``polarity'' vector that both guides and is affected by cell displacement. Aside
from the customary Hamiltonian  area constraint and cell--cell adhesion energy,
there is also a term for a  phenomenological motile force, implemented as a
``migration energy'' (dot product of the polarity vector and the cell centroid
position). The authors investigate how this motile force affects spatial
correlation of the velocity. They observe emergence of streaming patterns. The
authors also describe the influence of ``leader cells,'' whose polarity vector
is set to an external cue. Leaders are embedded in the tissue interior, and they
coordinating neighbors over some ``interaction distance.''

In a similar style of  experiment-model study, MDCK cells are seeded on a
circular adhesive domain in \cite{Doxzen2013}. Once a critical density is
attained, the confluent culture collectively rotates. The same behavior is then
captured in a CPM computational model that includes a motile force and a
polarity term that is reinforced by cell displacement, with some persistence
time. The authors show that rotation takes place when the size of the circular
domain is on the order of the correlation length of cells. That correlation
length, in turn, depends on the cell–-cell adhesion energy, the motile force
magnitude, and the polarity persistence time. Note the comparison with the purely
computational paper by \cite{Lober2015} using phase-field methods, where cell
collisions, rather than adhesive confluent cells, are the subject of focus. It is
also interesting to compare the CPM model in \cite{Doxzen2013} with the
vertex-based treatment of a similar situation in \cite{Lin2018}.

A recent paper, \cite{Jain2020}, demonstrates the fact that single cell
properties affect much of the collective behavior of a cell population. In
this ground breaking work, the authors link single cell Rac1 polarity to the
emergent rotations of confluent cells (from few to many) confined to a ring or
closed curve. A simple mechanical model captures the balance of directed
motility and contact forces to demonstrate the principles at work. This paper
spans subcellular to multicellular scales \LEK{and provides a great example of elegant experiments to be emulated by others. }

%%%%%%%%%%NEW TABLE%%%%%%%%%
\afterpage{%
\clearpage
\thispagestyle{empty}
\newgeometry{top=0.85in,left=0.75in,right=0.75in}
\centering
\begin{table}[!ht]
  \centering\scriptsize
  \begin{tabularx}{7in}{@{} t X t X @{}}
   \toprule
   \scriptsize\bf  \bf Bridging scales & \bf Levels of detail  & Refs. & \bf Experimental links \\
   \midrule

   \Signalling & {\bf Boolean} signaling network & \cite{Aracena2006} & Motivated by expts.\ \\

   \SingleCell & {\bf 3D phase-field} model & \cite{Winkler2019} & Motivated by expts.\ \\

   \SingleCell & {\bf 3D triangulated cells}  & \cite{Liedekerke2020} & Links to liver regeneration expts.\ \\

   \CollectiveCell & {\bf SPP}, pairwise forces, cell types & \cite{Carrillo2018} &  Pure theory, applied to expt'l design. \\

   \CollectiveCell & {\bf SPP},  Spherical cells, 2D, 3D; adhesion, repulsion, random forces & \cite{Stonko2015} & Compared to live images. \\

   \CollectiveCell & {\bf SPP}, 2D swarm, cell types,  differentiation & \cite{Volkening2015} & Compared to fish skin patterns. \\

   \CollectiveCell & {\bf vertex-based}, 3D & \cite{Osterfield2013}  & Integrated expt.-model \\

   \CollectiveCell & {\bf polygonal cells}, 2D, 3D, viscoelastic elements & \cite{Jamali2010} & Motivated by expts.\ \\

   \CollectiveCell & Cells as {\bf pairs of spheres} & \cite{Schnyder2017,Yang2020} & Motivated by \cite{Reffay2014}. \\

   \CollectiveCell & {\bf CPM} with cell polarity & \cite{Kabla2012,Londono2014} & Motivated by expts.\ \\

   \CollectiveCell & {\bf 3D vertex based model}, epithelia & \cite{hannezo2014} & Proposes new expts.\ \\

   \CollectiveCell & {\bf Spherical cluster};  Langevin eqn. &\cite{cai2016} & Integrated expt.-model \\

   \CollectiveCell & {\bf Continuum compressible fluid} tissue & \cite{Arciero2011} & Reproduces expts.\ \\

   \SingleCell, \CollectiveCell & {\bf Spherical cells}, detailed adhesion dynamics &  \cite{Frascoli2013} & Motivated by expts.\ \\

   \SingleCell, \CollectiveCell & {\bf SPP}, disk shaped cells, ECM, Cell--ECM interaction & \cite{Schluter2012} & Reproduces expts.\ of \cite{Lo2000}. \\

   \Signalling, \CollectiveCell & {\bf SPP} with alignment rules & \cite{Tarle2017} & Reproduces expts.\  \\

   \Signalling, \CollectiveCell & {\bf Spherical cells}, internal details & \cite{Drasdo1995,Drasdo2005} &  Reproduces expts.\  \\

   \Signalling, \CollectiveCell & {\bf Ellipsoidal cells} & \cite{Palsson2000,Palsson2008,Knutsdottir2017} & Motivated by expts.\ \\

   \Signalling, \CollectiveCell & {\bf Vertex based} & \cite{Lin2018,Lin2019} & Motivated by expts.\ \\

   \Signalling, \CollectiveCell & {\bf Vertex-based}; sub-cellular components  & \cite{Koride2018} & Motivated by expts.\ \\

   \Signalling, \CollectiveCell & {\bf Vertex-based}, intracellular signaling & \cite{Durney2018} & Motivated by expts.\ \\

   \Signalling, \CollectiveCell & {\bf FEM}, 1D approximation of 3D & \cite{Escribano2018} & Motivated by expts.\ \\

   \Signalling, \CollectiveCell & {\bf Cell agents}, repolarize after collisions &\cite{desai2013} & Integrated expt.-model. \\

   \Signalling, \CollectiveCell & {\bf SPP}, leaders,  followers, filopodia, chemical gradients & \cite{mclennan2012,mclennan2015,Schumacher2016} & Integrated expt.-model. \\

   \Signalling, \CollectiveCell & {\bf SPP} cells, polarity vectors, c-c coordination & \cite{Camley2017,Camley2018} &  -- \\

   \Signalling, \CollectiveCell & {\bf FEM} cells, signaling at cell edge, adhesion & \cite{Zhao2020} & Wound-healing, compared to expts. \\

   \Signalling, \CollectiveCell & {\bf Active media}, RDEs,  actomyosin, cell motion  &\cite{Banerjee2019} & -- \\

   \Signalling, \SingleCell, \CollectiveCell & {\bf 1D or 2D cells}, ODE signaling networks & \cite{Zmurchok2018,Bui2019} & Theoretical study. \\

   \Signalling, \SingleCell, \CollectiveCell & {\bf SPP}, 1D, rules for polarity coordination & \cite{George2017} & Motivated by expts.\ \\

   \Signalling, \SingleCell, \CollectiveCell &  {\bf deforming polygon} cells, signaling at nodes &\cite{Merchant2018} & Motivated by expts. \\

   \Signalling, \SingleCell, \CollectiveCell & {\bf CPM}, minimal intracellular signaling  & \cite{Rens2019} & Real cell traction forces \cite{Roux2016}. \\

   \Signalling, \SingleCell, \CollectiveCell & {\bf Phase-field} method, minimal intracellular signaling & \cite{Shao2010,Camley2013,Camley2014,Lober2015,Kulawiak2016} & Collision assays of \cite{Doxzen2013}. \\

    \bottomrule
  \end{tabularx}
  \caption{Summary of the modeling papers.
  Papers classified by 3 levels of organization: S, SCB, and CCB or tissue behavior.
  SPP models represent the cell shape statistically; common cell shape choices
  are spherical, ellipsoidal, or cylindrical cells.
  Models are categorized into: (1) CPM; (2) phase-field models;
  (3) vertex models; (4) particle models; (5) continuum models. CPM and phase-field models resolve cell shape well.
  c-c, cell-cell; CCB, collective cell behavior; CPM, Cellular Potts Models;
  expts., experiments; FEM, finite element methods; ODE, ordinary differential
  equations; RDE, reaction diffusion equation; S, signaling; SCB, single cell
  behavior; SPP, self-propelled particles.
  }\label{Table:ModelSummaryCombined}
\end{table}
\clearpage
\restoregeometry
}

%
% Section 3: Multi-scale computational modelling
%
\section{\LEK{Modeling recruited signaling networks and multiscale behavior in collective cell systems}}

At present, there are still
relatively few models that bridge from detailed underlying molecular mechanisms
through individual cell motility, all the way to tissue dynamics and collective
cell migration, though the number of such papers is growing.  There are 2
major issues that hamper such efforts. First, it is unclear how to deal with the
problem of combinatorial complexity in trying to understand numerous players and
interactions at each level. Decisions made at 1 stage affect other stages, and
exploring a multiplicity of assumptions is a challenge (but see
\cite{Manhart2018}). Second, complexity of the resulting models makes it
challenging to determine the range of possible behavior, let alone make sense
of overall principles and key components.

\LEK{Few papers specifically address the signaling modules that get recruited in
collective migration, beyond those that serve single cells migrating on their
own.  Here, we refer to signaling that is triggered by cell contact or
junctions, and that specifically affects the way that cells then interact.
Downstream responses might include changes in cell adhesion, migratory
potential, permissive or inhibitory control of cell division or apoptosis, or
relative sizes, polarity, or other aspects of cells. Input from the
environment in the form of mechanical tension or topography can influence
these signaling networks, promoting or inhibiting EMT.
Examples of this sort include some of the following.}

\LEK{The role of Merlin is highlighted in
\cite{Das2015}. Merlin, a negative regulator of Rac
\cite{Bretscher2002}, is a member of the
Ezrin-radixin-Moesin (ERM) family. When Merlin is bound to tight junctions between
cells, it inhibits Rac1, but when it is in the cytoplasm, it releases that
inhibition. At the same time, low Rac1 activity leads Merlin to become
stabilized at tight junctions
\cite{Das2015}. Hence, Merlin and Rac1 compete in a mutually inhibitory circuit,
and a balance between mechanical and chemical factors controls Merlin activity
and localization. The balance, in turn, regulates cell states that favor (high
Rac, low Merlin) or inhibit (high Merlin, low Rac) protrusion of cells at the front. Spatial
separation of Rac and Merlin activities can result in front--rear polarization in
collectively motile cells.}

\LEK{Mechanical cues and ECM topography appear to influence the adhesion and
migration of cells in an epithelial sheet. The signaling of YAP-Merlin-Trio
(YAP, Yes-associated protein) in
regulation of Rac and the expression of E-cadherin were investigated in
\cite{Park2019} on nanostructure ridge arrays that mimic ECM. A minimal model
for 2 signaling modules was proposed to account for the transition of YAP
activity with distance from the front edge of the sheet. These examples of how
cell--cell interactions depend on and influence both intracellular Rac gradients
and adhesion motivate future modeling efforts at the multiscale level.}

\LEK{An intriguing stepping stone on the route to the challenging eventual
targets are engineered tissues studied in synthetic biology. Since these are
designed with known components, they allow for more direct development of models
based on underlying mechanisms. We mention and example of this sort in what
follows.}

An interesting question is whether and how lessons from 1 level can be
simplified into rules for components at the next level and how to best include
the additional signaling pathways that get recruited at distinct levels of
organization.

Here, we mention a few representative examples but recognize that others may
exist of which we are as yet unaware.

\subsection{Bridging scales}

In many respects, we are currently seeing first steps in models that bridge
several layers of organization. Examples of this type
\Andreas{can be found in the work of Marzban and colleagues}
\cite{Marzban2018a,Marzban2018b}. Their model combines several modules: cell
polarization as in \cite{Maree2012}, a viscoelastic cytoskeleton, stress fiber
structure, cell motility as in \cite{Satulovsky2008}, and cell--substrate
interaction. The authors first model the polarization, motility, and durotaxis
in single cells, and then combine these with a cell--cell interaction module to
simulate the rotation of tens of cells in a confining 2D annulus. The paper
demonstrates a creative combination of a number of known cell representations to
bridge from subcellular properties to those of the collective.

\Andreas{%
In \cite{vanLiedekerke2019b}, the authors study tumor spheroid growth under
high mechanical compression. Their model represents cells as 3D
spheres. Commonly used cell contact models (e.g., the Hertz model)
do not take such large volume compression into account. The required corrections
to the contact models were calibrated using a high-resolution mechanical models
of cells \cite{Liedekerke2020}. We expect that similar approaches will in the
future improve the accuracy of coarse-grained tissue level simulations.}

\LEK{The intersection of cellular and developmental biology has provided
additional important examples of computational models that span multiple scales.
Some examples predate recent efforts by 2 decades, notably  computational
studies of cellular slime mold, {\it D.\ discoideum} morphogenesis, and
signaling in \cite{Savill1997,Maree2001}. Other examples such as
\cite{Hogeweg2000} are purely theoretical, aimed at exploring how random gene
circuits linked to cell adhesion, cell division, and some overall measure of
``fitness'' evolve into a zoological garden of multicellular structures. See
also \cite{Vroomans2015} for a recent work along a similar ``EvoDevo''
evolutionary developmental biology line. }

\subsection{Levels of detail}

By using very simple representations of subcellular events, \cite{Niculescu2015}
bridge from 1 cell to hundred(s) using CPM computations. The authors assumed an
elementary process that mimics, but does not explicitly depict, the effect of
actin on persistence of motion. Local protrusion is self-reinforcing, and
decays on some time scale. First showing how 2 simple computational parameters
tune the cell shape from keratocyte-like to amoeboid, they then simulate the
collective migration of a monolayer. The abbreviated
level of intracellular detail permits efficient computations. Note the
comparison with \cite{Marzban2018a}, where more costly FEM computations and
greater detail makes for greater computational cost and smaller number of cells
that can be readily simulated. Software packages as in \cite{Rubinacci2015} may
eventually make it more realistic to incorporate intracellular detail into
multiscale models.

\subsection{Links with experiments}

\LEK{In creating synthetic gene networks that regulate the adhesion protein
E-cadherin in real cells, Toda and colleagues \cite{Toda2018} succeeded to design
multicellular clusters that self-organize into distinct layers. The expression
of E-cadherin genes was placed downstream of cell-surface notch receptors. Notch
ligands on some cells activated notch receptors on neighbors, and in this way,
cell--cell interactions both influenced, and were influenced by intracellular
signaling.  The experiments were later linked to computational models in
\cite{Lam2020,Mulberry2020}. When details of the relatively ``simple'' genetic
circuits are known, as in such synthetic biology experiments, modelers can
bootstrap signaling models to learn how spatial influences and cell--cell
interactions shape the emergent tissue structures.   }

In \cite{Chauhan2011}, we find a link between intracellular signaling and tissue
morphogenesis. The authors show that the mutual antagonism of Rac and Rho can
affect the invagination and bending of epithelial that form a lens pit in the
eye development in a mouse.  Rho apparently controls the apical constriction of
cells via myosin, and Rac the elongation of those cells via F-actin assembly,
hence accounting for the conical angle formed by each cell and the overall
curvature of the tissue.

\LEK{Overall, more experimental papers that probe the circuits that get
recruited in the collective cell migration are needed. The papers
\cite{Das2015,Park2019} on the Merlin-Rac loop and on the link to YAP and
E-cadherin \cite{Park2019} should be followed up with more detailed
computational modeling and future rounds of experiments.}

\section{Discussion}

\begin{figure}[h]
  \centering
  \includegraphics[width=\textwidth]{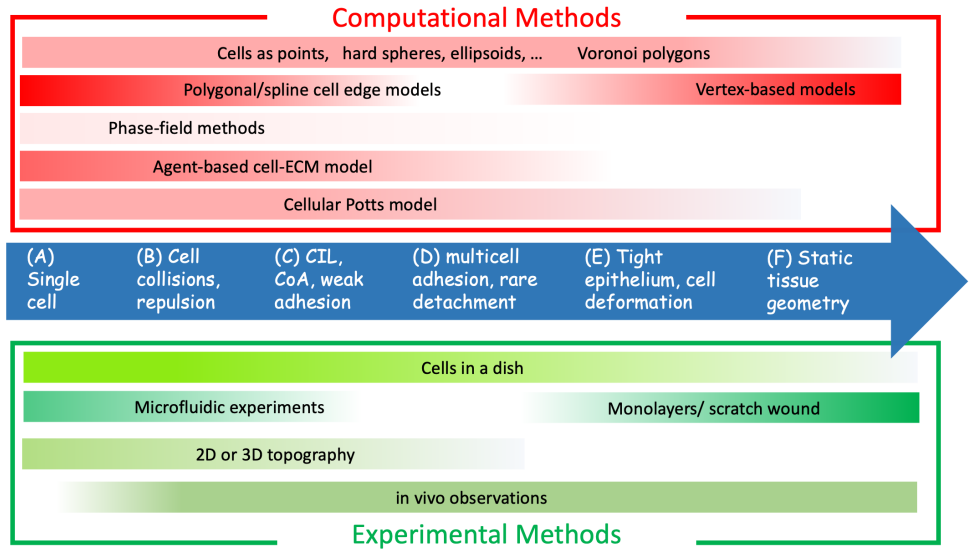}
\captionof{figure}[Summary of computational and experimental methods]{A summary of
  common modeling (top, red box) and experimental (bottom, green box) methods
  used to study cell behaviors from single cells to cell groups and up to
  tissues (left to right in increasing number of cells and increasing
  cell--cell adhesion).}\label{fig:SummaryModelExpt}
\end{figure}

In this paper, we described a small selection of modeling works on cell
migration that bridge from intracellular, to cellular and multicellular scales,
as shown in Fig~\ref{fig:BottomTop}. Many other excellent papers have been
omitted due to space limitations. That said, even from the fraction surveyed,
we find that a variety of computational and analytic methods are used at various
scales. Table~\ref{Table:ModelSummaryCombined}
organizes modeling papers by their subject and methodological approaches,
Table~\ref{Table:ExperimentSummary} classifies a few experimental papers by
their biological targets, and Fig~\ref{fig:SummaryModelExpt} summarizes the
ranges of relevance of both computational and experimental methods.
Our review has focused on the topic of single and collective cell migration and its regulation.
\LEK{Likely motivated by development of disease therapies and NIH funding, or
drug targets and support from pharmaceutical companies, the more medically
oriented subjects such as cancer, liver toxicity, or lung morphogenesis, have
fostered many generations of computational models. By comparison, the level of
basic scientific computational research on multiscale cell biology modeling is
still emerging.}

\begin{figure}[h]
  \centering
  \includegraphics[width=\textwidth]{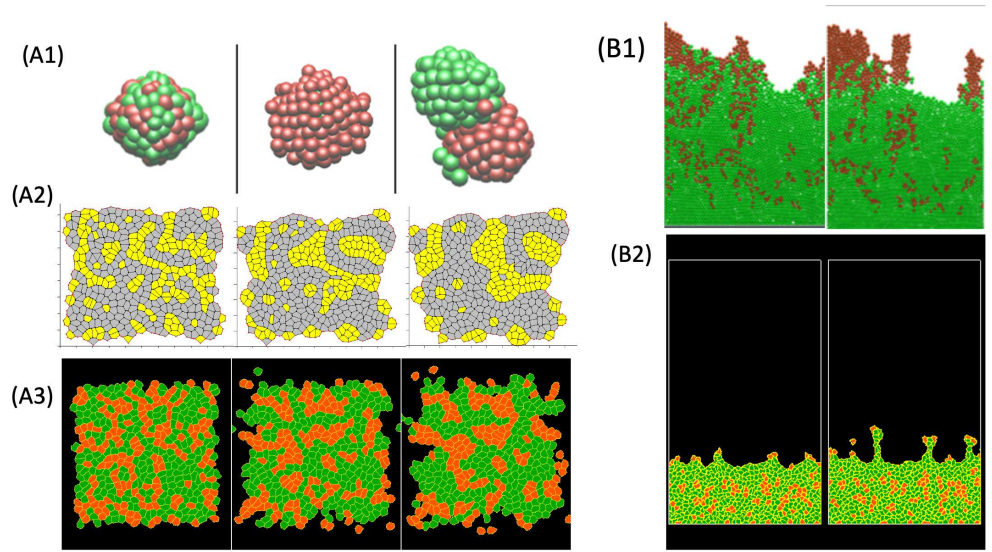}
  \captionof{figure}{Cell sorting (left) and wound-healing (right) captured by several
  distinct computational methods. (A1) Cells represented by deformable
  ellipsoids in 3D. Simulations by Hildur Knutsdottir based on code originally
  created by Eirikur Palsson (A2) Vertex-based simulations using CHASTE open
  source platform, run by Dhananjay Bhaskar. (A3) CompuCell3D cell-sorting
  simulations run by Dhananjay Bhaskar. In (A1--A3), there are 2 cell types
  with differing adhesion strengths to self and other cell type. (B1) and (B2)
  show deformable ellipsoids and CPM scratch wound models with 2 cell types.
  Cells with weaker adhesion (red) tend to segregate to the edge of the
  monolayer, acting as leader cells, and causing fingering of the front. (Compare with \cite{Yang2020}). Papers that compare distinct
  computational platforms include \cite{Osborne2017}.}
  \label{fig:CellSortingAndMono}
\end{figure}

We have seen that some computational techniques that  work well at the
single-cell level, become too costly or excessive at the tissue level. (See also
review of computational models in
\cite{Danuser2013,Yang2019,Camley2017b,Alert2019,Osborne2017} and
\cite{Rejniak2011} for cancer.) We also encountered topics where diverse
computational techniques lead to similar predictions. (See
Fig~\ref{fig:CellSortingAndMono}, and \cite{Osborne2017} for comparisons.)
\LEK{We still find instances where 1 or another group claims that their
computational method of choice outperforms others or has fewer unrealistic
features. In many cases, such claims are, at best, unfortunate and overlook
essential shared attributes. In other cases, they skip over relative advantages
versus disadvantages of the distinct schemes. We believe that the field needs
more rational comparisons of how custom-build computations perform against a
host of ``benchmark'' test problems, as for example, shown in
\cite{Osborne2017}, and/or deeper comparative analysis of cell-surface mechanics
models that demonstrate equivalence of distinct approaches, as in
\cite{Magno2015}. }

\LEK{While more journals now require that simulation codes be made publicly
available, in practice, this is only a half measure. Many researchers, and most
biologists do not have the  software packages or expertise to read, execute, and
run codes in different formats. We recommend that, moving forward, the
community should invest more directly in standardizing open-source software,
with several specific aims: (1) ease of operation, user friendliness, and rapid
learning curve; (2) basic built-in code for most major computations, including
reaction--diffusion solvers, particle position and collision solvers, cell shape
and cell--cell adhesion, intra and intercellular signaling, and so on; (3) ease
of sharing ``code,'' as, for example, in the simple small Morpheus \texttt{xml}
files --- these preserve exact details of each simulation run; and (4) ease of
development of new plug-ins to allow the capabilities to expand with the needs
of the community. We recommend that groups invest resources in helping to
develop such shared platforms and that funding bodies (NIH, NSF, and NSERC) make it
a priority to support such developments. Certainly, at initial stages, this
development requires teams of computational experts who can establish solid,
robust platforms, and create training manuals or instructional videos to recruit
new users. Morpheus, a software system developed at the Technische
Universit\"{a}t Dresden, is 1 example along these lines, but others are
needed.}

Returning to questions posed in the Introduction, we can draw a few general
conclusions.

\subsection{Bridging scales}

Building from the bottom up, models that include detailed
molecular interaction or gene networks (such as the Boolean models in
\cite{Hetmanski2016,Aracena2006}) encode many details and occasionally reveal
attractors or identify missing components \cite{Aracena2006}. They are harder to
understand on their own.  Detailed models such as \cite{Ditlev2009} benefit from
insights of preceding more basic models \cite{Mogilner2002}. At lower levels of
detail but still arguably ``bottom up'' are studies that show how certain
properties of molecular components and interactions result in specific cell
behavior such as polarization
\cite{Otsuji2007,Jilkine2007,Maree2006,Maree2012}. Other work starts from
observed cell dynamics to infer the likely signaling pathways at play
\cite{Lin2012,Lin2015,Park2017} or likely underlying mechanisms
\cite{Lacayo2007,Manhart2018}, that we could denote ``top down.'' The latter,
\cite{Manhart2018}, also demonstrates the idea that it is preferable to weed
through and discuss many models, particularly some that fail, with reasons for
that failure, not merely aiming for a single optimal model. This kind of process
helps to build a greater intuitive understanding of the specific key mechanisms
required to explain observed phenotypes.

\LEK{We recommend that modelers avoid publishing large-scale models as a ``fait
accompli,'' so as to ``be the first to accurately simulate'' some entire
comprehensive behavior of choice. Instead, we suggest that modelers should
provide due description of the steps taken in developing such models, with all
informative failures and successes. This would help the community to assimilate
the insights that resulted from an entire program of research. }

\LEK{In a related, but separate vein, the field is in need of rational consensus
and standard practices for stepping between size scales and levels of
organization. To draw a common analogy, our ability to easily use and appreciate
global geography as well as local structure has been transformed by the way that
Google maps seamlessly allows us to zoom down to street-level views and back out
to the globe as a whole. The algorithms that reveal or blur over specific
details as we step up or down were logically constructed for easy navigation
and optimal viewing of many levels of complexity. Nothing like this currently
exists in the realm of cell biology. The fact that structures and interactions
change on a rapid timescale make this a tough issue, to be sure, but one
deserving more attention.}

Associating rule of behavior with a specific hierarchy is possible once we have
sufficient familiarity with the biology and predictions of basic models. This
can help to bridge hierarchies and avoid the fog of complexity. Mathematical
methods such as dynamical systems, PDEs, and bifurcation analysis can help to
find and account for emergent properties and universal principles in such basic
models \cite{Kholodenko2006}. This is one of the strengths of the mathematical
tools. A weakness is that these methods currently work well for small systems of
differential equations, but not for large and complex systems.

Once we understand the repertoire of a single cell, we can move up to 2,
3, or many cells. Experimental observations of small cell groups provide
good opportunities for understanding how to bridge from single to collective
behavior.  Modeling can then also explore the advantages of group migration,
chemotaxis \cite{Camley2017,Camley2018}, durotaxis \cite{Escribano2018}, etc.,
in larger groups. For a large enough number of cells, condensing the details
into simpler rules becomes expedient. For example, while polarity is represented
by PDEs and patterns inside a single cell, it can then be simplified, depicted
by a direction vector \cite{George2017,Kabla2012} in place of a full internal
gradient of Rho or Rac for multiple cells.

\subsection{Level of detail}
\LEK{The current computing power at our
disposal allows increasingly detailed models to be constructed, with tens or
even hundreds of components, and many more parameters. The temptation to create
such a model and to present it as a mechanistic representation of real cells is
hard to resist. Yet, before going down this very intricate route, important
questions should be considered: What do we expect to learn? Might we have left
out something important? Have we included superfluous detail that obscures the
basic structure? Are we certain that our parameter regime faithfully corresponds
with true rates and quantities? As previously noted, even though we can
construct highly complex models, our comprehension of that detail is limited.}

In general, best practices seem to invoke back and forth attempts to include
important aspects, simplify to understand, throw out secondary factors, and go
back to including detail.  Simplifications lead to insights and rigorous
analytic results but leave large gaps to real biology. However, returning to
detailed versions of a model once the major insights are at hand (or vice versa)
is helpful. While \cite{Maree2006} explains polarization in a motile cell, for
example, \cite{Mori2008} extracts some of the key properties and interactions
that guarantee that it could work in a mathematically tractable mini-model.

Furthermore, ``toy models'' consisting of small sets of ODEs or PDEs can help
to formulate universal principles that work across many biological examples and
many scales. Here, we can mention the concepts that mutual inhibition or positive
feedback, which result in bistability and hysteresis. Simple ``modules''
consisting of specific types of interacting components, whether molecules,
cells, or animals have prototypical behaviors as switches, oscillators, or
other dynamics, and, with molecular diffusion or random motion, engender
patterns or waves \cite{Kholodenko2006,Tyson2003}.

\subsection{Model-experiment links}
Biology is an experimental science at its core,
and models that illuminate its mysteries must eventually meet and concur with
biological evidence. Modern methods such as machine learning can help match a
vast dataset with the best candidate model(s) as shown elegantly by
\cite{Manhart2018}. At the same time, modelers, physicists, and biophysicists
can make real contributions by using their craft and theories to unlock
important facts that are not at all apparent otherwise \cite{Pollard2016}.

\LEK{Several of our examples demonstrate that some experimental papers have
explicitly addressed signaling pathways that mediate cell--cell communication in
collective cell behavior.  As pointed out eloquently by a reviewer of this
paper, ``experiments can only tell [us ... what are] upstream and downstream
regulators. We need mathematical models to incorporate this information with the
intracellular networks to form a signaling network on a multicellular level to
study how a group of cells processes signals collaboratively.''}

\bigskip\noindent
{\bf Acknowledgements: }
We are grateful to the Pacific Institute for Mathematical
Sciences for providing space and resources for AB’s postdoctoral
research.  We are grateful to former and current members of the Keshet-Feng
Research Group for helpful discussions over many years that solidified our
appreciation of this area of research. We thank Hildur Knutsdottir and Dhananjay
Bhaskar for permission to use results shown in
Fig~\ref{fig:CellSortingAndMono} that they produced while they were UBC
students.

\nolinenumbers
% Use the PLoS provided BiBTeX style
\bibliographystyle{plos2015}
% \bibliography{bibliography.bib}

%%%%%%%%%%%%%%%%%%%%%
%%%%%%.  END %%%%%%%%%%%%
%%%%%%%%%%%%%%%%%%%%

\end{document}